\documentclass[prb,preprint,eqsecnum,showpacs]{revtex4}
\usepackage{amsmath}
\usepackage{graphicx}
\usepackage{amssymb} 
\usepackage{amsfonts}

\begin{document}

\title{Interpolation of the Josephson interaction in highly
  anisotropic superconductors from a solution of the two dimensional
  sine-Gordon equation. }
       
\author{Yadin Y. Goldschmidt}
\author{Sandeep Tyagi}

\affiliation{Department of Physics and Astronomy, University of Pittsburgh,
Pittsburgh, Pennsylvania 15260}

%\maketitle

\begin{abstract}
In this paper we solve numerically the two dimensional elliptic
sine-Gordon equation with appropriate boundary conditions. These
boundary conditions are chosen to correspond to
the Josephson interaction between two adjacent pancakes belonging to
the same flux-line in a highly anisotropic superconductor. An
extrapolation is obtained between the regimes of low and high
separation of the pancakes. The resulting formula is a better
candidate for use in numerical simulations than previously derived
formulas. 
\end{abstract}

\maketitle
\section{Introduction}

High-temperature superconductors belong to the class of superconducting
materials known as type II that allow
for partial magnetic flux penetration whenever the external field satisfies
$H_{c1}<H<H_{c2}$ \cite{tinkham,blatter,brandt}. The flux penetrates
the sample in the form of flux-lines (FL's), each containing a quantum unit
$\phi_0=hc/2e$ of flux. At low temperature the FL's form an ordered
hexagonal lattice (Abrikosov lattice) due to their their mutual
repulsion. At high temperature and/or magnetic field this lattice
melts due to thermal fluctuations \cite{exp1,exp2,cubitt,exp3,exp4}.

High-temperature superconductors are anisotropic material which are
made from stacks of superconducting layers associated with CuO$_2$
planes. The layers are weekly coupled to each other. The parameter
measuring the anisotropy is $\gamma$, defined as $\gamma^2=m_z/m_\perp$,
where $m_z$ and $m_\perp$ denote 
the effective masses of electrons moving along the $c$ axis
(perpendicular to the superconducting planes) and the
$ab$ plane, respectively. While
for the material YBa$_2$Cu$_3$O$_{7-\delta}$ known as YBCO the
anisotropy is somewhere between 5-7, for the material
Bi$_2$Sr$_2$CaCu$_2$O$_{8+\delta}$ known as BSCCO, the anisotropy is
estimated to be between 10 to a 100 times larger. 

For BSCCO and highly anisotropic materials similar to it, each FL is
represented more faithfully by a collection of objects referred to as
``pancakes''. Pancakes are centered at the superconducting planes. Each pancake
interacts with every other pancake, both in the same plane and in
different planes. The
interaction can be shown to consist of two parts. The first part is
called the electromagnetic interaction (or simply magnetic) and it
exists even in the case that 
the layers of the materials are completely decoupled, so no current
can flow along the $c$-axis of the sample. A pancake vortex located in one
plane gives rise to screening currents in the same plane as well
as in all other planes. A second pancake vortex, located elsewhere,
interacts  with the screening currents induced by the first
pancake \cite{artemenko}. This interaction has been calculated by Clem 
and others \cite{clem}. Two pancakes in the same plane interact with a
repulsive interaction while pancakes in different planes attract one another.

The second part of the interaction among pancake vortices is the
so-called Josephson interaction \cite{artemenko,clem1,blatter}.
It results from the fact that there is a
Josephson current flowing between two superconductors separated by an
insulator and this current is proportional to the sine of the phase
difference of the superconducting wave functions. The superconductors
in the present case are the different CuO$_2$ planes. When two
pancakes belonging to the same stack and residing in adjacent
planes move away from each other, the phase difference that originates
causes a Josephson current to begin flowing between the planes. This
results in an attractive interaction between pancakes that for
distances small compared to $r_g\equiv \gamma d$ is approximately
quadratic \cite{artemenko,blatter} in the distance. When the two adjacent
pancakes are separated by a distance larger than $r_g$, a ``Josephson
string'' is formed, whose energy is proportional to its length
\cite{clem1}.

When doing simulations of flux lattices it is important to include
correctly the strength of the interaction among pancakes. It turns out
that for anisotropies smaller than about 150, the Josephson
interaction is dominant over the magnetic interaction and the later
can be included only in the form of an effective in-plane interaction
\cite{blatter,ryu,st_yyg}. For higher anisotropies the magnetic 
interaction must be included among all pairs of pancakes , but
the Josephson interaction still matters for any finite anisotropy 
\cite{st_yyg}. The form of the Josephson
interaction used in simulation was first introduced by Ryu {\it et al.}
\cite{ryu} based on the Lawrence-Doniach model \cite{ld}. These authors
used a certain approximation which can be somewhat improved. It is the
aim of this paper to review the approximations made for the Josephson
interaction and to suggest a better approximation to be used in
numerical simulations. In order for future Monte Carlo simulations to
yield a better agreement with experimental results it is important to
choose the form of the interaction among pancakes to be as close as
possible to the exact interaction in real materials. In the
simulations the interaction among pancakes belonging to different FL's
is dominantly electromagnetic for the range of magnetic fields used in
experiments in the vicinity of the melting transition. For the same
FL, the Josephson interaction is present predominantly for nearest
neighbor pancakes which are displaced from an alignment along the
z-axis. For pancakes separated by more than
one plane the electromagnetic interaction is present. If a
kink is present in more than one place along the same FL the total interaction 
is given to a good approximation as a sum of the pair interactions,
provided the FL does not deviate too much from a straight line which
is usually the case on the solid side of the melting line.

\section{The model}
Our starting point is the Lawrence-Doniach \cite{ld,blatter} Gibbs free-energy
functional, 
\begin{eqnarray}
\mathcal{G}[\psi _{n},\mathbf{a}] &=&\int d^{2}\mathbf{R} \ d
\,\sum_{n}\left[\alpha |\psi
_{n}|^{2}+\frac{\beta }{2}|\psi _{n}|^{4}+\frac{\hbar ^{2}}{2m}\left\vert
\left( \frac{\mathbf{\nabla }^{(2)}}{i}+\frac{2\pi }{\phi _{0}}\mathbf{a}%
^{(2)}\right) \psi _{n}\right\vert ^{2}  \right. \notag \\
&&\left.+\frac{\hbar ^{2}}{2Md^{2}}\left\vert \psi _{n+1}\exp \left( \frac{2\pi i}{%
\phi _{0}}\int_{n\,d}^{(n+1)d}dz\,a_{z}\right) -\psi _{n}\right\vert ^{2}\right] 
\notag \\
&&+\int d^{2}\mathbf{R} dz\left( \frac{b^{2}}{8\pi }-\frac{\mathbf{b}\cdot 
\mathbf{H}}{4\pi }\right) ,
\label{eq:ldfe}
\end{eqnarray}
where $\psi _{n}~$represents the superconducting order parameter in the $%
n^{th}$ CuO$_2$ layer,
$\mathbf{a}^{(2)}$ is the vector potential in the plane, and $a_z$ its
component perpendicular to the planes. $\mathbf{b}$ is the
local magnetic field and $\mathbf{H}$ is the externally applied
field. $\alpha$ and $\beta$ are the Ginzburg-Landau parameters. $m$ and
$M$ are the effective masses along the $x-y$ and $z$ directions
respectively. $d$ is the separation between the superconducting planes.
$\phi_0$ is the flux quantum. The
integration along the $z$-direction has been replaced by a discrete
summation over the superconducting layers.
In the London approximation the absolute value of $\psi_n$ is treated as
a constant and its phase $\phi_n$ is varying. The Gibbs functional
becomes (dropping some constants):
\begin{eqnarray}
\mathcal{G}=\int d^{2}\mathbf{R} \,\frac{\varepsilon _{0}d}{2\pi }\left\{ 
\sum_{n}\left( \mathbf{\nabla }^{(2)}\phi
_{n}+\frac{2\pi }{\phi _{0}}\mathbf{a}^{(2)}\right) ^{2} \notag
\right. &&\\
\left. +\frac{2}{\gamma^2 d^{2}}\sum_{n}
\left[ 1-\cos \left( \phi _{n+1}-\phi _{n}+\frac{%
2\pi }{\phi _{0}}\int_{nd}^{(n+1)d}dz\,a_{z}\right) \right]%
\right\} \notag \\ 
+\int d^{2}\mathbf{R}dz\left( \frac{b^{2}}{8\pi }-\frac{\mathbf{b\cdot
      H}}{4\pi}\right), 
\label{eq:ldmodel}
\end{eqnarray}
where
\begin{equation}
\varepsilon _{0}= 2\pi \frac{\hbar ^{2}|\psi _{n}|^{2}}{2m}=\frac{\phi
  _{0}^{2}}{(4\pi \lambda )^{2}}, 
\label{eq:ensc}
\end{equation} 
where $\lambda$ is the penetration depth and and $\gamma=\sqrt{M/m}$
is the anisotropy. Minimization with respect to $\mathbf{a}^{(2)}$ gives
\begin{equation}
\lambda ^{2}\mathbf{\triangle a}^{(2)}=d\sum_{n}\delta(z-n\,d)\left[ 
\mathbf{a}^{(2)}+\frac{\phi _{0}}{2\pi }\mathbf{\nabla }^{(2)}\phi _{n}%
\right] ,  \label{eq:2cvp}
\end{equation}
where $\mathbf{\triangle }$ is the 3-dimensional Laplacian.
Minimization with respect to $a_{z}$ gives
\begin{equation}
\triangle a_z = \frac{4 \pi}{c} j_J \sin (\Phi _{(n,n+1)}),
\label{eq:zcvp}
\end{equation}
where 
\begin{equation}
\Phi_{n+1,n}=\phi _{n+1}-\phi _{n}+\frac{2\pi }{\phi _{0}}%
\int_{nd}^{(n+1)d}dz~a_{z}
\label{eq:gephase}
\end{equation}
is the gauge invariant phase difference between the layers $n$ and
$n+1$, and 
\begin{eqnarray}
  j_J=\frac{c\phi_0}{8\pi^2\lambda^2\gamma^2 d}
\label{eq:jcurrent}
\end{eqnarray}
is the Josephson-coupling current density between layers. Minimization
with respect to $\phi _{n}$ gives
\begin{equation}
 \Delta^{(2)}\phi _{n}+\frac{2\pi }{\phi _{0}}%
\mathbf{\nabla }^{(2)}\cdot \mathbf{a}^{^{(2)}} =\frac{1}{\gamma
  ^{2} d^2}\left[\sin (\Phi_{n,n-1})-\sin (\Phi_{n+1,n})\right].
\label{eq:efphi}
\end{equation} 
We have used the Coulomb gauge $\mathbf{\nabla} \cdot \mathbf{a}=0$.
Eqs. (\ref{eq:2cvp}), (\ref{eq:zcvp}) and (\ref{eq:efphi}) are to be
solved with the appropriate boundary conditions and the solution must
be substituted back into the expression for the Gibbs
free-energy.

An isolated pancake residing in plane $n$ is a singular solution of the
equation for the phase of the wave function which -n the limit of 
$\mathbf{R} \rightarrow \mathbf{R}_n$ satisfies
\begin{equation}
\mathbf{\nabla }^{(2)}\phi _{n}(\mathbf{R })=-\frac{\mathbf{\hat{z}}\times
\left({\mathbf{R }}-{\mathbf{R}}_{n}\right)}{\left({\mathbf{R
    }}-{\mathbf{R}}_{n}\right)^2 }, 
\label{eq:pancake}
\end{equation}
where $\mathbf{R}_n$ denotes the center of the pancake and
$\mathbf{\hat{z}}$ is a unit vector in the z-direction. This solution
becomes exact in the whole plane in the infinite anisotropy limit. As one
fully encircles the pancake the phase $\phi_n$ 
changes by $2\pi$, and is singular at the center of the pancake. In
the case of infinite anisotropy, the full solution of 
Eqs. (\ref{eq:2cvp}), (\ref{eq:zcvp}) and (\ref{eq:efphi})
can be found \cite{clem} and from it one can deduce the interaction
energy for two pancakes in the same plane or in different planes. 
But for more complicated configurations, or when the anisotropy is
finite, the solution can only be found approximately.

Consider Eq.~(\ref{eq:efphi}) for the $(n+1)$-layer and for the $n$'th layer
respectively. Subtracting the second case from the first we obtain
\begin{eqnarray}
 \Delta^{(2)}(\phi_{n+1}- \phi _{n})+\frac{2\pi }{\phi _{0}}%
(\mathbf{\nabla }^{(2)}\cdot \mathbf{a}^{(2)}(nd+d)-
\mathbf{\nabla }^{(2)}\cdot \mathbf{a}^{(2)}(nd))\notag \\
 =\frac{1}{\gamma ^{2} d^2}\left[2\sin (\Phi_{n+1,n})-\sin (\Phi_{n+2,n+1})-
\sin (\Phi_{n,n-1})\right].
\label{eq:ff1}
\end{eqnarray}
 adding and subtracting the term $(2 \pi d/\phi_0)\triangle a_z$ it becomes
\begin{eqnarray}
 \Delta^{(2)}(\phi_{n+1}-\phi _{n}+\frac{2\pi}{\phi_0}d a_z) 
+\frac{2\pi }{\phi _{0}}
(-d \triangle^{(2)}a_z+d\partial_z\mathbf{\nabla }^{(2)}\cdot \mathbf{a}^{(2)}(nd))\notag \\
 =\frac{1}{\gamma ^{2} d^2}\left[2\sin (\Phi_{n+1,n})-\sin (\Phi_{n+2,n+1})-
\sin (\Phi_{n,n-1})\right].
\label{eq:ff2}
\end{eqnarray}
using the coulomb gauge we see that
\begin{eqnarray}
-\triangle^{(2)}a_z+\partial_z\mathbf{\nabla }^{(2)}\cdot
\mathbf{a}^{(2)}=-(\triangle^{(2)}+\partial_z^2)a_z=-\triangle a_z.
\label{eq:vpm}
\end{eqnarray}
We are now in a position to simplify Eq.~(\ref{eq:ff2}) by using
Eq.~(\ref{eq:gephase}) and Eq.~(\ref{eq:zcvp}) together with
Eq.~(\ref{eq:jcurrent})  to
finally find \cite{artemenko2,bulaevskii}
\begin{eqnarray}
  \triangle^{(2)}\Phi_{n+1,n} = \frac{1}{\gamma ^{2} d^2}\left[2\sin 
(\Phi_{n+1,n})-\sin (\Phi_{n+2,n+1})-\sin (\Phi_{n,n-1})\right]\notag \\
+\frac{1}{\gamma^2\lambda^2}\sin(\Phi_{n+1,n}).
\label{eq:geeq}
\end{eqnarray}
which amazingly involves only the gauge invariant phase difference
among the layers. So far this equation is nearly exact. The problem is
that it represents an infinite set of coupled equations.

\begin{figure}
\centering
\includegraphics[width=1.5in,height=2.5in]{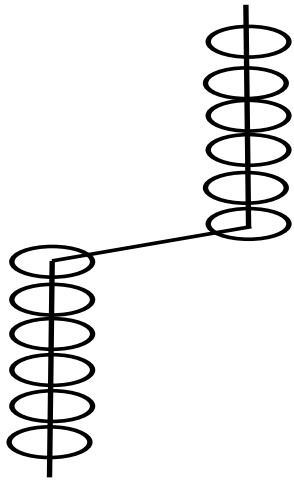}
\caption{A kink in a configuration of pancakes aligned along the
  $z$-direction.} 
\label{fig:kink}
\end{figure}

Consider now a FL which consists of a stack of pancakes. Assume that 
there is a kink in the FL in the sense that
the $n$'th and the $(n+1)$'th pancakes are not on top of each other but
are shifted a distance $w$ along the $x$ direction. The pancakes with
index $\ge n+1$ are all aligned along the $z$-direction as well as the
pancakes with index $\le n$. See Figure (\ref{fig:kink}). In this situation
one might think that except for $\Phi_{n+1,n}$ all the phase
differences vanishes. However this is not entirely true since because
of the Josephson coupling there is induced phase differences away from
the location of the kink. For example in the situation that $w$ is
very large a so called Josephson vortex is formed. Although its core
only extends a distance $d$ in the $z$-direction and a distance $\gamma d$ in
the $y$-direction, the associated magnetic field extends a
distance $\lambda$ in the $z$-direction and $\gamma \lambda$ in the
$y$-direction.

\begin{figure}
\centering
\includegraphics[width=3in,height=3in]{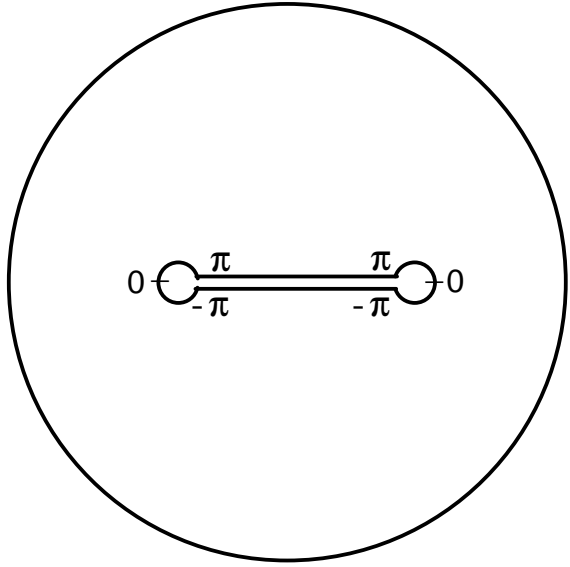}
\caption{Boundaries and boundary conditions for the solution of the two-dimensional elliptic
  sine-Gordon equation. } 
\label{fig:bc}
\end{figure}

\begin{figure}
\centering
\includegraphics[width=2in,height=2in]{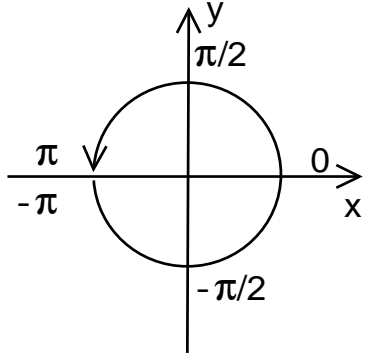}
\caption{definition of atan2 }  
\label{fig:atan2}
\end{figure}

Nevertheless in order to truncate the infinite set of equations we
will assume that we can neglect all the gauge invariant phases
as compared to $\Phi_{n+1,n}$. This turns out to be quite satisfactory for
distances much smaller than $\gamma \lambda$. Later we will add an
overall factor to partially compensate for the approximation made.
If we make this approximation, and further neglect the last term on
the right hand side of Eq.~(\ref{eq:geeq}) since $\lambda^2 \gg d^2$, we
obtain \cite{feigelman}
\begin{eqnarray}
\mathbf{\Delta }^{(2)}\Phi=\frac{2}{\Lambda^2}\sin(\Phi),  
\label{eq:SG}
\end{eqnarray}
which is the well-known elliptic sine-Gordon equation in two
dimensions, where we put $\Phi \equiv \Phi_{n+1,n}$ and $\Lambda \equiv
\gamma d$. The energy associated 
with a solution to this equation due to the Josephson currents is
found by evaluating the expression
\begin{eqnarray}
  \mathcal{G}_J = \frac{d \varepsilon_0}{\pi \Lambda^2} \int
  d^2\mathbf{R}(1-\cos(\Phi(\mathbf{R})).
\label{eq:jenergy}
\end{eqnarray}
The boundary conditions for the sine-Gordon equation appropriate to
the configuration under discussion are fixed on the geometry depicted
in Fig.~(\ref{fig:bc}). The outer boundary is a large circle of radius
$R_m$ which tends to infinity. On this boundary we set $\Phi = 0$. 
The inner boundary consists of two tiny circles the centers of which
are a distance $w$ apart, connected by a
thin rectangle. On the inner boundary we take the solution to satisfy 
\begin{eqnarray}
  \Phi(\mathbf{R})= {\rm atan2}(y,\ x-w/2) - {\rm atan2}(y,\ x+w/2),
\label{eq:datan}  
\end{eqnarray}
where $\mathbf{R}=(x,y)$ and atan2 is the 4-quadrant inverse tangent
whose values lie in the interval $(-\pi,\ \pi)$ as depicted  in
Fig.~(\ref{fig:atan2}). It is to be distinguished from ${\arctan}(y/x)$
whose value ie in the interval $(-\pi/2, \pi/2)$. 
Thus the values of $\Phi$ is given approximately
by the values depicted in Fig.~(\ref{fig:bc}). These values become
exact as the radius of the inner circles and the width of the
rectangle tend of zero. 

We proceed by first discussing the solution under certain limiting
conditions. Under these conditions we can come up with analytic
expressions. One such limit is the case $w \ll \Lambda$. Under this
condition it is a good approximation to linearize the sine-Gordon
equation and solve instead the equation
\begin{eqnarray}
  \label{eq:linsg}
\mathbf{\Delta }^{(2)}\Phi=\frac{2}{\Lambda^2}\Phi,   
\end{eqnarray}
which has a solution
\begin{eqnarray}
  \label{eq:sollin}
  \Phi(\mathbf{R})=\frac{\sqrt{2}w}{\Lambda}\sin(\theta) K_1\left(\frac{\sqrt{2}R}{\Lambda}\right),
\end{eqnarray}
where $(R,\theta)$ are the polar coordinates of the vector $\mathbf{R}$
 and $K_1$ is the modified Bessel function of order 1. With the given
 boundary condition this solution is only valid for $R \gg w$. We see
 that for $w \ll R \ll \Lambda$ the solution becomes
 \begin{eqnarray}
   \label{eq:rless}
   \Phi(\mathbf{R})\sim \frac{w}{R}\sin(\theta),\ \ \ \ \ w \ll R \ll \Lambda,
 \end{eqnarray}
whereas for $R \gg \Lambda$ we obtain
\begin{eqnarray}
  \label{eq:rmore}
  \Phi(\mathbf{R}) \sim
  \sqrt{\frac{\pi}{\sqrt{2}}}\frac{w}{\sqrt{R\Lambda}}
\sin(\theta)\exp\left(-\frac{\sqrt{2}R}{\Lambda}\right),\ \ \ R \gg \Lambda.
\end{eqnarray}
To obtain the energy we note that since $\Phi$ is small we can expand
the cosine in Eq.~(\ref{eq:jenergy}) to second order in $\Phi$ and
substituting from Eq.~(\ref{eq:rless}) cutting the integration at $w$ for
lower limit and $\Lambda$ for an upper limit we obtain for the
potential energy as a function of $w$
\begin{eqnarray}
  \label{eq:potenergys}
  V(w) \approx \frac{\epsilon_0 d w^2}{2 \pi \Lambda^2}\int_w^\Lambda
  \frac{dR}{R}\int_0^{2\pi}d\theta\ \sin^2(\theta)=\frac{\epsilon_0
    d}{2}\left(\frac{w}{\Lambda}\right)^2
  \ln\left(\frac{\Lambda}{w}\right)+ O\left(\left(\frac{w}{\Lambda}\right)^2\right). 
\end{eqnarray}
The reader may wonder how the normalization of the solution in
Eq.~(\ref{eq:sollin}) has been determined. This is due to the boundary
conditions on the inner boundary. In the region $w \ll R \ll \Lambda$
the term on the right hand side of Eq.~(\ref{eq:linsg}) can be dropped and
$\Phi$ satisfies Laplace's equation. A solution that satisfies the
boundary conditions can be written in the first quadrant in terms of
the ordinary arctan function as follows:
\begin{eqnarray}
  \Phi(\mathbf{R})= \arctan\left(\frac{y}{x-w/2}\right) - \arctan\left(\frac{y}{x+w/2}\right),
\label{eq:atan}  
\end{eqnarray}
which for $x \gg w$ gives
\begin{eqnarray}
  \label{eq:rless2}
  \Phi(\mathbf{R})\approx \frac{yw}{x^2+y^2}=\frac{w}{R}\sin(\theta),
\end{eqnarray}
in agreement with Eq.~(\ref{eq:rless}). The solution given in
Eq.~(\ref{eq:sollin}) matches correctly to the solution of the Laplace's
equation satisfying the correct boundary conditions and hence it is
properly normalized.  

\begin{figure}
\centering
\includegraphics[width=4in,height=3in]{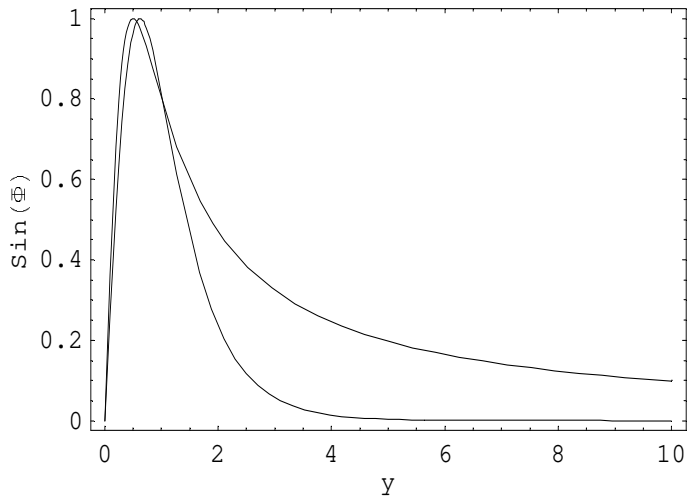}
\caption{Comparison of the solution given by Eq.~(\ref{eq:cch}) (longer
  tail) to the solution given by Eq.~(\ref{eq:sinkink}) plotted
  against $y$ in units for which $\Lambda=1$ and $\lambda_c=100$.}  
\label{fig:compare1}
\end{figure}

The other asymptotic limit that can be done analytically is the limit
$w \rightarrow \infty$. In that limit the solution should be
independent of $x$, and thus $\Phi$ should satisfy the one-dimensional
sine-Gordon equation
\begin{eqnarray}
  \label{eq:1dSG}
\frac{\partial^2}{\partial y^2}\Phi(y)=\frac{2}{\Lambda^2}\sin(\Phi(y)),  
\end{eqnarray}
with the solution in the upper half plane satisfying 
\begin{equation}
  \Phi(0^+)=\pi,\ \ \ \Phi(\infty)=0,
\end{equation}
and on the lower plane
\begin{equation}
  \Phi(0^-)=-\pi,\ \ \ \Phi(\infty)=0.
\end{equation}
A well-known kink solution to the sine-Gorgon equation is given by 
\begin{eqnarray}
  \label{eq:kink}
  \Phi(y)= {\rm sign}(y) \times 4 \arctan(\exp(-{|y|\sqrt{2}}/{\Lambda})),
\end{eqnarray}
which implies
\begin{eqnarray}
\label{eq:sinkink}
\sin(\Phi(y))=2 \sinh(y\sqrt{2}/\Lambda)/\cosh^2(y\sqrt{2}/\Lambda).  
\end{eqnarray}
The energy of this solution per unit length in the $x$ direction is
given by
\begin{eqnarray}
  \label{eq:kinken}
   \frac{d \varepsilon_0}{\pi \Lambda^2} \int_{-\infty}^\infty dy\ 
  \left(1-\cos\left(4
      \arctan(\exp(-{|y|\sqrt{2}}/{\Lambda}))\right)\right)= \frac{4
    \epsilon_0 d}{\sqrt{2}\pi\Lambda}.
\end{eqnarray}
If $w$ is large but finite, we can take the total length of the
``string'' as $w$, and to a good approximation 
\begin{eqnarray}
  V(w)= \frac{4
    \epsilon_0 d}{\sqrt{2}\pi} \left(\frac{w}{\Lambda}\right),\ \ \ \ \ \
  w \gg\Lambda,
\label{eq:potenergyl}
\end{eqnarray}
which is linear in the separation between the two singularities
(pancakes). In this regime the non-linearity of the sine-Gordon
equation is crucial.
In the case of $w\rightarrow\infty$, which can be referred to as a
single Josephson vortex, it is possible to find an
approximate solution directly to the infinite set of equations given by
Eq.~(\ref{eq:geeq}). The solution found by Clem, Coffey and Hao
\cite{cc,clem1} reads
\begin{eqnarray}
  \label{eq:cch}
  \sin(\Phi_{n+m+1,n+m}(y))= \frac{\Lambda}{ \lambda_c^2}\frac{y K_1(r)}{r}.
\end{eqnarray}
with
\begin{eqnarray}
  \label{eq:r}
  r=\frac{\Lambda}{2\lambda_c}\left(1+4y^2/\Lambda^2+4 m^2\right)^{1/2}.
\end{eqnarray} 
where we put
\begin{eqnarray}
  \label{eq:lambdac}
  \lambda_c\equiv \gamma \lambda,
\end{eqnarray}
The value $m=0$ corresponds to 
the position of the kink (Josephson
vortex). Note that in the $z$-direction the phase difference decreases
rapidly but not abruptly. Also there is an additional length scale
$\lambda_c \gg \Lambda$ not present in the approximation made above. If we plot the
rhs of Eq.~(\ref{eq:cch}) for $m=0$ on the same graph as $\sin(\Phi)$ from
Eq.~(\ref{eq:sinkink}) for say $\Lambda=1$ and $\lambda_c=100$ we see, as
depicted in Fig.~(\ref{fig:compare1}), that
both rise linearly in $y$ up to a maximum for $y\sim \Lambda$ and then
both decay exponentially, on scales $\lambda_c$ and $\Lambda$ respectively.
When evaluating the energy with the solution given by
Eq.~(\ref{eq:cch}) one finds instead of the result represented by the
rhs of Eq.~(\ref{eq:kinken}) a value of
\begin{eqnarray}
  \label{eq:kinken2}
  \left(\ln\left(\frac{\lambda_c}{\Lambda}\right)+1.12\right)\frac{\epsilon_0d}{\Lambda},
\end{eqnarray}
per unit length of the vortex in the $x$-direction. Using a different
approximation Koshelev \cite{koshelev} claims that the constant in the
last equation should be 1.55 instead of 1.12. In any case the
appearance of $\ln(\lambda/d)$ is the result of the fact that the
magnetic field decays over a distance of $\lambda$ and not $d$ in the
$z$-direction as was implied by the neglecting the phase differences
away from the $\Phi_{n+1,n}$. So we can correct this coefficient by
introducing the logarithm by hand in front of the solution of the
sine-Gordon equation as will be discussed further below.

\section{A numerical solution}

The main problem is how to interpolate the potential energy of two
pancakes as their separation changes from $w \ll \Lambda$ to $w \gg
\Lambda$. 
In this case an analytical solution is not available and one
has to seek a numerical solution. This we achieved using the
finite-element method implemented by the PDE tool
of the program MATLAB. We solve the sine-Gordon equation \ref{eq:SG},
in the plane with the boundary condition displayed in
Fig.~(\ref{fig:bc}) and on the inner boundary given more accurately
by Eq.~(\ref{eq:datan}).  
We have chosen $\Lambda=1$. The outer radius has been taken to be $R_{max}=15$
for $0.2 \le w \le 10$, and $R_{max}=5$ for $w < 0.2$. The radius of
the inner circles has been taken to be $R_{min}=0.04$ for $1.2 \le w
\le 10$,
$R_{min}=0.02$ for $0.6 \le w \le 1.2$ and $R_{min}= w/30$ for  $w
<0.6$. 
The width of the rectangle has always been taken to be $R_{min}/2$.

\begin{figure}
\centering
\includegraphics[width=4 in,height=4in]{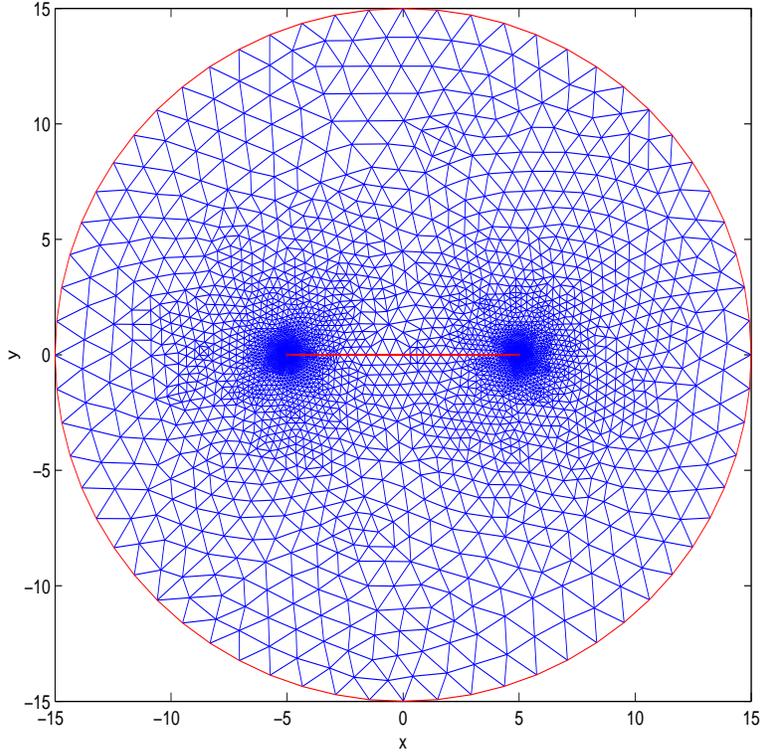}
\caption{Triangular grid for the case $w=10$. Distances are in units
  of $\Lambda$.} 
\label{fig:gridw10}
\end{figure}

\begin{figure}
\centering
\includegraphics[width=4in,height=4in]{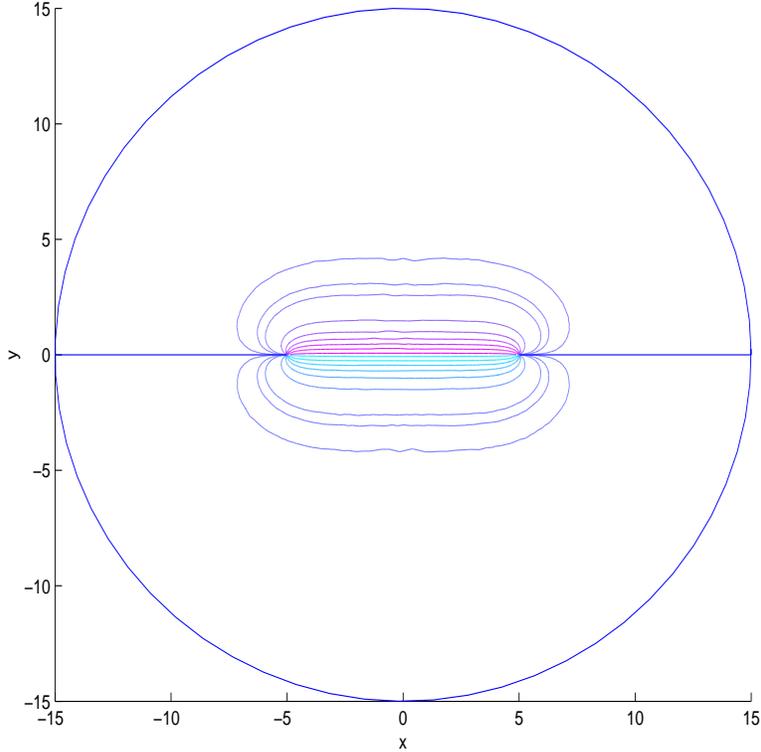}
\caption{Contour plot of the solution of the sine-Gordon equation for
  $w=10$.  Contour values from bottom to top: 0(boundary), -0.01,
  -0.05, -0.1, -0.5, -1, -1.5, -2, -2.5, -3, 3, 2.5, 2, 1.5, 1, 0.5,
  0.1, 0.05, 0.01, 0 (boundary).
}  
\label{fig:solw10}
\end{figure}

In Figure \ref{fig:gridw10} we show example of the triangular grid used for
the case $w=10$ and the corresponding solution $\Phi(x,y)$ is depicted
in Figure \ref{fig:solw10}. This figure depicts contour lines of
equal $\Phi$ for some specified values as indicated in the figure
caption. It can be verified that the contour lines 
for the range of $x$ in the middle between the singularities ($-3<x<3$), are 
almost identical to those obtained from the sine-Gordon kink solution
given by Eq.~(\ref{eq:kink}) which is valid for $w \rightarrow
\infty$. For this
solution the contour lines are always parallel to the $x$-axis.


\begin{figure}
\centering
\includegraphics[width=5in,height=4in]{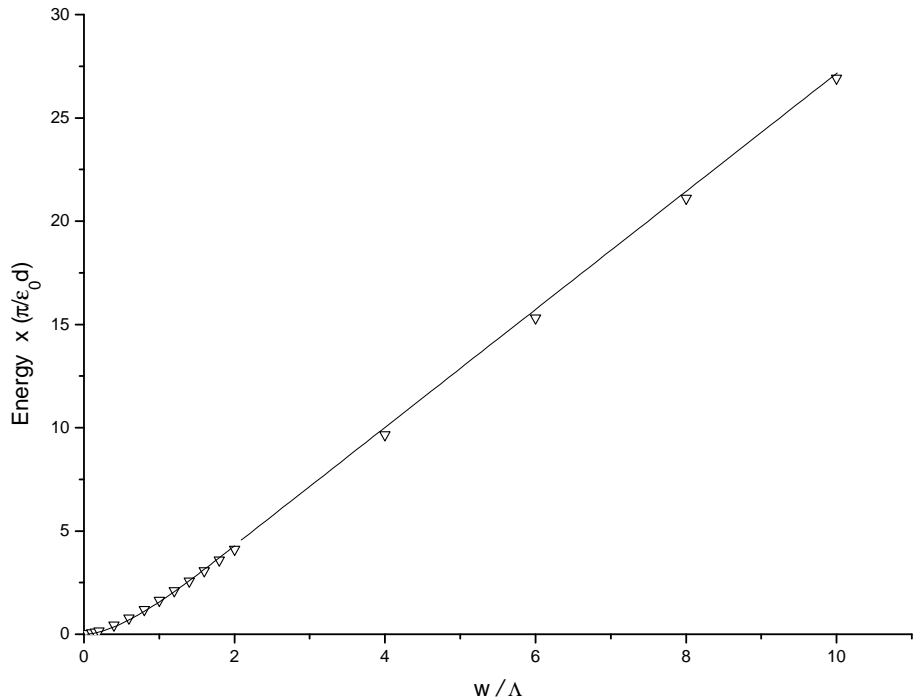} 
\caption{The energy as obtained from the numerical solution of the
  sine-Gordon equation (inverted triangles) and a fit (solid line) as
  given by Eq.~(\ref{eq:energyfit}).}
\label{fig:enplot}
\end{figure}

The energy versus pancake separation $w/\Lambda$ is plotted in Figure \ref{fig:enplot}.
A good fit to the energy plot is as follows:
\begin{eqnarray}
 V &=& \left(\frac{\epsilon_0 d}{\pi}\right) 0.707 \left(\frac{w}{\Lambda}\right)^2
 \ln\left(\frac{9 \Lambda}{w}\right), \ \ \ \ \ \ \ \ w \le 2\Lambda \notag\\
   &=& \left(\frac{\epsilon_0 d}{\pi}\right) \left( 2.828
     \left(\frac{w}{\Lambda}\right) -1.414\right), \ \ \ \ \ \ \ 2\Lambda <
   w,
\label{eq:energyfit}
\end{eqnarray}
Notice that the factor $2.828=4/\sqrt{2}$ agrees with the coefficient
of the analytical solution given by Eq.~(\ref{eq:potenergyl}). 
The prefactor of the quadratic term is smaller than that given by
Eq.~(\ref{eq:potenergys})since we made a fit up to a value of $w$ larger
than $\Lambda$ in order to obtain a simpler extrapolation formula. The
constant inside the logarithm captures the quadratic term on the right of
Eq.~(\ref{eq:potenergys}) which is $O((w/\Lambda)^2)$ and whose
coefficient is not determined by the simple argument given.
( For very small values of $w$ a good fit is given by
 $(\pi V/\epsilon_0 d) = 1.57\ (w/\Lambda)^2\ \ln(2.6 \Lambda/w)$ ). 
In order to compensate for the approximation made by the use of the
sine-Gordon equation, we will now rescale the energy by a factor that
will reproduce correctly
the Josephson vortex energy as given by Eq.~(\ref{eq:kinken2}) which
takes the exponential decay along the $z$-direction into account. Thus
we suggest the following expression for the Josephson energy among two
adjacent pancakes belonging to the same FL and displaced horizontally
by an amount $w$:
\begin{eqnarray}
  \label{eq:jospot}
  V_{new}(w)  &=& \epsilon_0 d\ (1+\ln(\lambda/d))\ 0.25 \left({w}/{\Lambda}\right)^2
 \ln\left({9 \Lambda}/{w}\right), \ \ \ \ \ w \le 2\Lambda \notag\\
   &=&  \epsilon_0 d\ (1+\ln(\lambda/d))\ \left(
     \left({w}/{\Lambda}\right) - 0.5\right), \ \ \ \ \ \ \ \ \ \ \ \ \ \ \ 2\Lambda <
   w.
\end{eqnarray}
This expression should be compared with the extrapolation proposed by
Ryu {\it et al.} \cite{ryu}, that reads, after correcting for a
misprint by a factor of $2/\pi$ in that reference, and shifting by a constant
\begin{eqnarray}
  \label{eq:ryu}
  V_{ryu}(w)  &=& \epsilon_0 d\ (1+\ln(\lambda/d))\ 0.25
  \left({w}/{\Lambda}\right)^2,
 \ \ \ \ \ \ \ \ \ \ \ \ \ \ \ \ \ \    w \le 2\Lambda \notag\\
   &=&  \epsilon_0 d\ (1+\ln(\lambda/d))\ \left(
     \left({w}/{\Lambda}\right) - 1\right), \ \ \ \ \ \ \ \ \ \ \ \ \ \ \ \ \ 2\Lambda <
   w.
\end{eqnarray}

These authors did not take into account the logarithmic dependence
$\ln(\Lambda/w)$ of the potential as given by
Eq.~(\ref{eq:potenergys}) and kept only the quadratic dependence on
$w$. They adjusted the coefficient to obtain a match of the function
and its first derivative at the matching point. Note that our formula
Eq.~(\ref{eq:jospot}) also has the feature of a continuous first
derivative (related to the force between pancakes) at the matching point.  
For BSCCO the value of $\ln(\lambda/d)\approx 4.7$. Thus the additive
factor of $1$ in the coefficient is rather small compared to the logarithm
and this value was used by Ryu {\it et al.} instead of the more
precise (but still approximate) value $1.12$ derived in
Ref.~\onlinecite{clem1}. Koshelev \cite{koshelev}, using a different 
approximation, claims that this value should actually be 1.55 and one
might consider using this value instead of the additive 1 in the
suggested new formula.
 
\begin{figure}
\centering
\includegraphics[width=3in,height=3in]{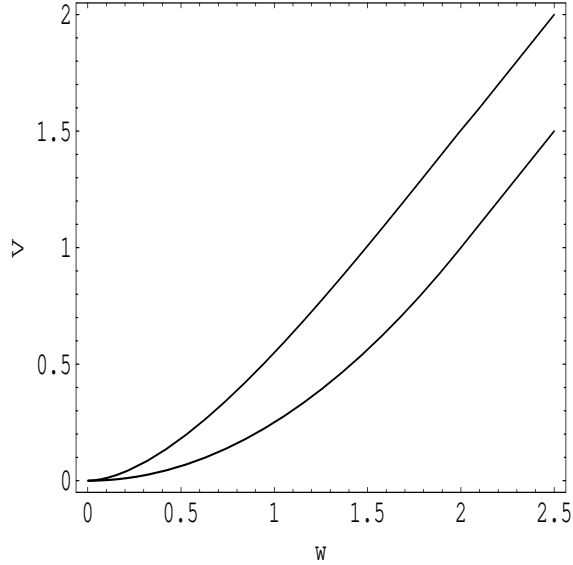}
\caption{Comparison of the two formulas given by Eq.(\ref{eq:jospot}) and
  Eq.~(\ref{eq:ryu}) for $V_{new}$ (upper curve) and $V_{ryu}$ (lower
  curve) with the prefactor omitted. $\Lambda$ was taken to be equal
  to 1.}
\label{fig:compot}
\end{figure}

In Figure \ref{fig:compot} we show a comparison of the two
formulas given by Eq.~(\ref{eq:jospot}) and Eq.~(\ref{eq:ryu}), 
where the coefficients $\epsilon_0 d (1+\ln(\lambda/d))$ have been
omitted. The new interpolation should provide a better fit to the true
potential in real systems. In the next section we discuss the results
of simulations we have carried out with the new formula versus the old
one.

\section{Numerical simulations using the new interpolation}

We have carried out numerical simulations using the new formula for
the Josephson interaction between pancakes. We also included the
electromagnetic interaction along the lines discussed in our previous
publication \cite{st_yyg}. This simulation has been done with 36
planes and 36 pancakes per plane. The temperature dependence of the
penetration depth used for the results displayed below is 
$\lambda^2(0)/\lambda^2(T)=1-T/T_c$ with $\lambda(0)=1700$ {\AA},
$T_c=90$ K and $d=15$ {\AA} . 
The simulation method used is a multilevel Monte Carlo, as discussed
in our previous publications \cite{st_gold,st_yyg}, with the only
difference being the formula used for the Josephson interaction.

\begin{figure}
\centering
\includegraphics[width=6in,height=3in]{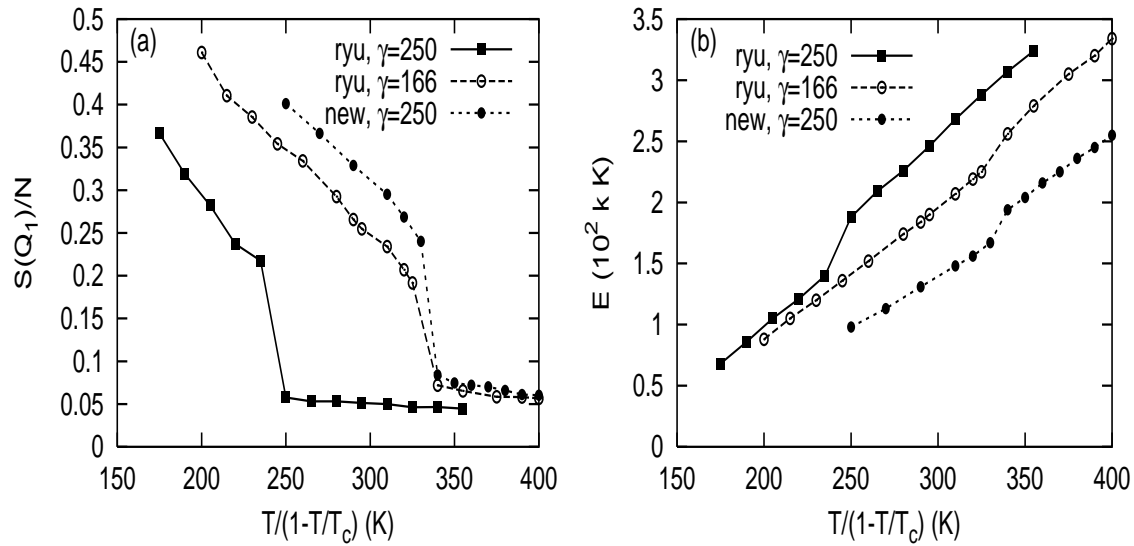}
\caption{Comparison of the results of multilevel Monte Carlo simulations using the new
  formula for the Josephson interaction as compared to the same simulations
  using the old formula derived by Ryu {\it et al.} for a
  magnetic field equal 100 Gauss. The
  electromagnetic interaction is also included. The structure factor
  (a) and the energy (b) are displayed. }
\label{fig:simul}
\end{figure}

In Figure \ref{fig:simul} we show the
results of a simulation for BSCCO using the parameter $\gamma=250$ 
for the anisotropy and $B=100$ Gauss for the applied field (curves
labeled as 'new'). This is to
be compared with the results using the old formula for values of
$\gamma=166$ and $\gamma=250$ (curves labeled as 'ryu'). It should be
emphasized that the figures refer to as 'ryu' are not results of Ryu
et. al \cite{ryu} but rather our simulations using their formula (corrected by
a factor $2/\pi$) for the Josephson interaction. The figures show the 
structure factor and the energy as a function of temperature. The
absolute value of the energy is not meaningful since a constant has
been subtracted, only energy differences are meaningful. 

We see that roughly speaking the
location of the melting transition for $\gamma=250$ using the new
formula occurs at the same temperature as for $\gamma=166$ with the
old formula, although some details like the magnitude of the structure
factor and the magnitude of the energy jump are not identical so it
not just a trivial shift. The melting transition with the new 
formula appears a little sharper. But roughly speaking there 
is a scaling of the anisotropy by a factor $\sim 1.5$ compared with
using the old formula. Thus our previous estimates for the anisotropy of
experimental samples should be roughly increased by this factor. 
In Ref.~\onlinecite{st_yyg} we estimated the anisotropy of
experimental samples used in the experiments of Majer {\it et
al.}\cite{majer} to be in the range 250-450 depending on the actual 
temperature dependence of the penetration depth. This value should
thus be increased to the range 375-675 with an average of 525. 
This is in good agreement with recent measurements that places the
anisotropy in experimental samples to be about 550  \cite{gaifullin}.
All the other qualitative results
of our previous simulations \cite{st_yyg} should remain roughly the 
same apart from the fact the curves should be relabeled to correspond 
to a higher anisotropy.

\section{Conclusions}

In this paper we derived a new approximate formula for the Josephson
interaction between neighboring pancakes belonging to the same
flux-line. The formula extrapolates better between the limiting
analytic solutions of the sine-Gordon equation than the previously
derived formula by Ryu {\it et al.} \cite{ryu}. We have seen that
we get agreement with the experimental results of Majer {\it et al.}
\cite{majer} for a value of the anisotropy $\sim 525$ in good agreement with
recent measurements of Gaifullin {\it et al.} \cite{gaifullin} who estimate
the anisotropy of their sample to be about $550$.

There are still many features of BSCCO system in the presence of
disorder that are not fully understood like the properties of the
different phases of the vortex system found in the case of point
disorder \cite{fuchs}. Future simulations to reproduce the different phases
should make use of the improved formula for the Josephson interaction
in order to get realistic results which are in good agreement with
experimental measurements.

\section{Acknowledgments}
This work is supported by the US Department of Energy (DOE), Grant 
No. DE-FG02-98ER45686.

\end{document}